# Isochoric thermal conductivity of solid nitrogen

V. A. Konstantinov, V. G. Manzhelii, V. P. Revyakin, and V. V. Sagan

Institute for Low Temperature Physics & Engineering of NASU, 61103, 47 Lenin Ave.,
Kharkov, Ukraine.

The isochoric thermal conductivity of solid nitrogen has been investigated on four samples of different densities in the temperature interval from 20 K to the onset of melting. In $\alpha$-$N_2$ the isochoric thermal conductivity exhibits a dependence weaker than $\Lambda \propto 1/T$; in $\beta$-$N_2$ it increases slightly with temperature. The experimental results are discussed within a model in which the heat is transported by low-frequency phonons or by «diffusive» modes above the mobility boundary. The growth of the thermal conductivity in $\beta$-$N_2$ is attributed to the decreasing «rotational» component of the total thermal resistance, which occurs as the rotational correlations between the neighboring molecules become weaker.

**Introduction**

The thermal conductivity of simple molecular crystals is determined by both translational and orientational motion of molecules in the lattice sites. This motion can be either oscillatory or rotational depending on the relation between the noncentral force and the rotational kinetic energy. Except for rare cases (quantum crystals), the motion of molecules at rather low temperatures is inherently oscillatory: at T=0 the molecules execute orientational vibrations about equilibrium directions. As the temperature rises, the r.m.s. amplitudes of the librations increase and the molecules can jump over some accessible orientations. This may lead to a phase transition because the long-range orientation order disappears. By choosing crystals with different parameters of molecular interaction and varying the temperature, it is possible to change the degree of the orientational order and investigate the effect of the molecule rotation upon the thermal conductivity.

Owing to their rather simple and largely similar physical [1-2], the $N_2$ – type crystals ($N_2$, CO, $N_2O$ и $CO_2$) consisting of linear molecules come as suitable objects for such studies. In these crystals the noncentral part of the molecular interaction is determined mostly by the quadrupole force. At low temperatures and pressures, these crystals have a cubic lattice with four molecules per unit cell. The axes of the molecules are along the body diagonals of cube. In $N_2$ and $CO_2$ having equivalent diagonal directions the crystal symmetry is *Pa3*, for the noncentrosymmetrical molecules CO and $N_2O$ the crystal symmetry is *P2$_1$3*.

In $CO_2$ and $N_2O$ the noncentral interaction is very strong and the long-range orientational order can persist up to their melting temperatures. In $N_2$ and CO the barriers impeding the rotation of the molecules are an order of magnitude lower; as a result, orientational disordering phase transitions occur at 35,7 and 68,13 K, respectively. In the high-temperature phases, the $N_2$ and CO molecules occupy the sites of the HCP lattice of the spatial group *P6$_3$/mmc*.

For a correct comparison with theory, thermal conductivity must be measured at constant density, which excludes the thermal expansion effect. Such investigations were made on $CO_2$ and $N_2O$ in [3]. Significant deviations from the dependence $\Lambda \propto 1/T$ were observed at $T \geq \Theta_D$. It was shown that these departures occurred when the thermal conductivity was approaching its lower limit. The concept of the lower limit of thermal conductivity [4] is based on the following: the mean free paths of the oscillatory modes participating in heat transfer are essentially limited, and the site-to-site heat transport proceeds as a diffusive process.

The goal of this study was to investigate the isochoric thermal conductivity of solid nitrogen in both orientationally ordered and orientationally disordered phases. Earlier, the thermal conductivity of nitrogen was investigated only under saturated vapour pressure [5-7].

**Experimental technique.**

Constant-volume investigations are possible for molecular solids having a comparatively low thermal pressure coefficient $\left(\dfrac{\partial P}{\partial T}\right)_V$. Using a high-pressure cell, it is possible to grow a solid sample of sufficient density. In subsequent experiments it can be cooled with practically unchanged volume, while the pressure in the cell decreases slowly. In samples of moderate densities the pressure drops to zero at a certain characteristic temperature $T_0$ and the isochoric condition is then broken; on further cooling, the sample can separate from the walls of the cell. In the case of a fixed volume, melting occurs in a certain temperature interval and its onset shifts towards higher temperatures as density of samples increases. This is seen, for example, in the V-T phase diagram of solid $N_2$ [2] in Fig. 1. The deviations from the constant volume caused by the thermal and elastic deformation of the measuring cell were usually no more than 0.3% and could be taken into account.

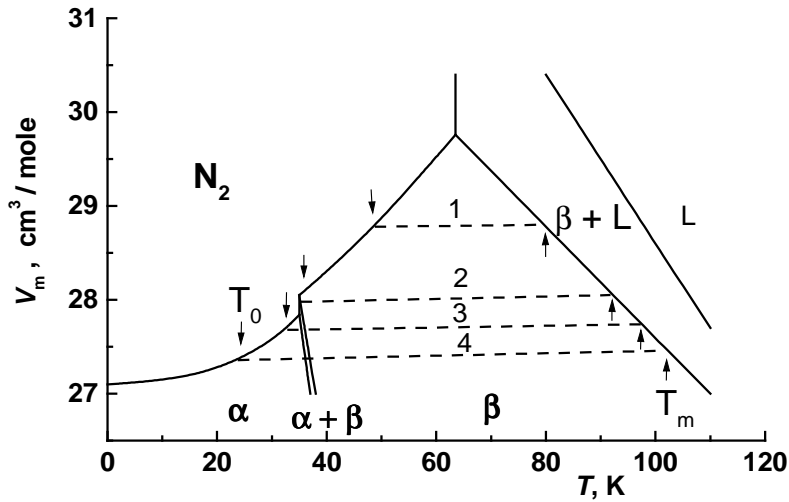

Fig.1. V-T phase diagram of solid $N_2$ according to [1,2]: (----)-molar volumes; arrows show the onset of V=const condition and melting.

The investigation was made using a steady-state technique in a coaxial-geometry setup. The measuring beryllium bronze cell was 160 mm long with the inner diameter 17.6 mm. The maximum permissible pressure in it was 800 MPa. The inner measuring cylinder was 10.2 mm in diameter. Temperature sensors (platinum resistance thermometers) were placed in special channels of the inner and outer cylinders to keep them unaffected by high pressure. In the process of growing the temperature gradient over the measuring cell was 1-2 K/cm. The pressure in the inflow capillary was varied within 50-200 MPa to grow samples of different densities. When the growth was completed, the capillary was blocked by freezing it with liquid hydrogen, and the samples were annealed at premelting temperatures for one to two hours to remove density gradients. After measurement the samples were evaporated into a thin-wall vessel and their masses were measured by weighing. The molar volumes of the samples were estimated from the known volume of the measuring cell and the sample masses. The total (dominant) systematic error of measurement was no more than 4% for the thermal conductivity and 0.2% for the volume. The purity of $N_2$ was no worse than 99.98%.

**Results and discussion**

The isochoric thermal conductivity of solid $N_2$ was investigated on four samples of different densities in the temperature interval from 20 K to the onset of melting. The experimental thermal conductivities are shown in Fig. 2. With solid lines for smoothed values and a dashed line for measurement under saturated vapor pressure [2, 5-7]. Under the same P, T- conditions, the discrepancy between our and literature data was no more than 4%. The molar volumes $V_m$, temperatures $T_0$ (onset of V=const condition) and $T_m$ (onset of sample melting) are shown in Table 1. This information is also available in Fig. 1.

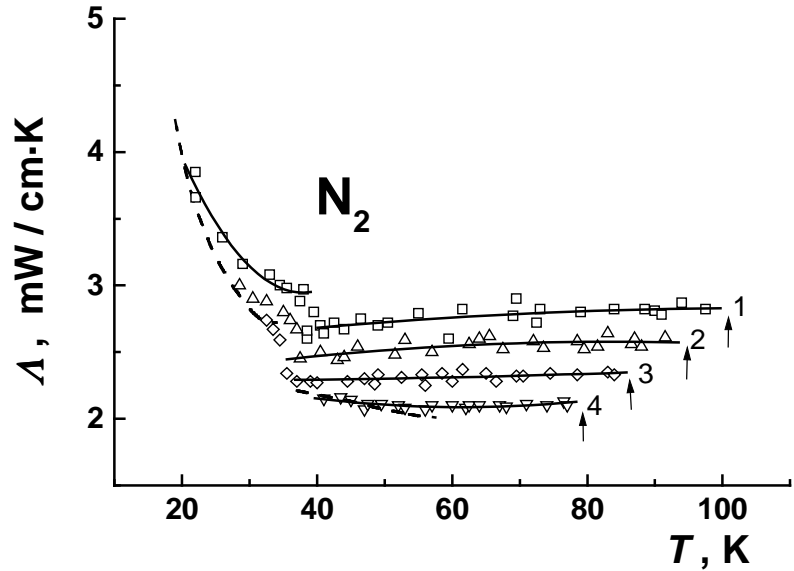

Fig.2. Isochoric thermal conductivity of four solid $N_2$ samples of different densities (see Table.1): (——)-smoothed values, (----)-measurement under saturated vapor pressure according to [2, 5-7], arrows show the onset of melting.

Table 1. Molar volumes $V_m$, temperatures $T_0$ (onset of V=const condition) and temperatures $T_m$ (onset of melting).

| Sample № | $V_m$ cm$^3$/mole | $T_0$ K | $T_m$ K |
|---|---|---|---|
| 1 | 27.36 | 24 | 102 |
| 2 | 27.68 | 33 | 96 |
| 3 | 27.98 | 36 | 92 |
| 4 | 28.76 | 48 | 80 |

In $\alpha$-$N_2$ the temperature dependence of the isochoric thermal conductivity is weaker than $\Lambda \propto 1/T$ and in similar to that observed for $CO_2$ and $N_2O$ [3]. The thermal conductivity is practically constant immediately before the $\alpha \rightarrow \beta$ transition. Earlier, the thermal conductivity was observed to grow in orientationally disordered phases of some mo-lecular crystals [8]. The Bridgman coefficients $g = -(\partial \ln \Lambda/\partial \ln V)_T$ calculated from the experimental results are $6.0 \pm 0.8$ for $\alpha$-$N_2$ at T=35 K and $4.3 \pm 0.5$ for $\beta$-$N_2$ at T=60 K.

The orientational motion of the molecules in $\alpha$-$N_2$ manifests itself as large-angle librations (immediately before the $\alpha \rightarrow \beta$ transition the r.m.s. libration amplitudes $\langle\vartheta^2\rangle^{1/2}$ exceed 30°) attended with hopping over a limited set of equivalent orientations related by the group symmetry elements [1]. The frequency of reorientations approaches $10^{-11}$ sec$^{-1}$ near the $\alpha \rightarrow \beta$ transition [9]. The analysis of heat capacity data suggests that practically free precession of the molecules is observed in $\beta$-$N_2$ after the phase transition, which is accompanied by axial vibration through an angle $\theta$ with respect to the hexagonal axis of the cell [1]. In the $\beta$-phase of $N_2$ the frequency of reorientation's varies from $9.5 \times 10^{-11}$ sec$^{-1}$ immediately after $\alpha \rightarrow \beta$ transition to $5.5 \times 10^{-12}$ sec$^{-1}$ before melting [9]. This exceed considerably the Debye frequency $1.0 \times 10^{-12}$ sec$^{-1}$. No distinct libration modes were detected in $\beta$-$N_2$ in inelastic neutron scattering experiments, and even the observed translational acoustic phonons were broadened considerably due to the translation-orientation interaction, excluding the case of the smallest wave vectors [10].

Remembering that the orientational motion of the molecules in $\alpha$-$N_2$ is essentially librational in character, the thermal conductivity can be calculated within a model in which the heat is transferred by low-frequency phonons or «diffusive» modes above the mobility boundary. Earlier, the model was used to calculate the thermal conductivity of $CO_2$ and $N_2O$ [11].

Let us describe the thermal conductivity as:

$$\Lambda(T) = 3nk_B v \left(\frac{T}{\Theta_D}\right)^3 \int_0^{\Theta_D/T} l(x) \frac{x^4 e^x}{(e^x - 1)^2} dx \qquad (1)$$

where $\Theta_D = v(\hbar/k_B)(6\pi^2 n)^{1/3}$, n is the number of atoms (molecules) per unit volume, v is the polarization-averaged sound velocity, l(x) is the phonon mean free path. At $T \geq \Theta_D$ the mean free path is mainly determined by the umklapp processes $l(x) = l_u$ where:

$$l_u = \frac{\lambda^2}{CT} \quad , \quad C = (12\pi^3/\sqrt{2}) \, n^{-1/3} \, (\gamma^2 k_B / mv^2) \qquad (2)$$

Here $\lambda$ is the phonon wavelenth, $\gamma$ is the Gruneisen constant, m is the atomic (molecular) mass, C is the numerical coefficient. In the first approximation, the translation-orientation interaction in molecular crystals leads to extra scattering which can be taken into account through simple renormalization of the coefficient C [12]. Remembering that the smallest phonon mean free path is about half the wavelength: $l(x) = \alpha\lambda/2$, where $\alpha \approx 1$, the «diffusivity» boundary $\lambda^*$ can be found as:

$$\lambda^* = \frac{\alpha CT}{2} \qquad (3)$$

which corresponds to the effective temperature $\Theta_* = 2hv/\alpha k_B CT$. (It is assumed that $\Theta_* \leq \Theta_D$, otherwise $\Theta_* = \Theta_D$). Below, the term «diffusive» is applied to the modes whose mean free paths reached the smallest values [11]. The integral of thermal conductivity is subdivided into two parts describing the contributions to the thermal conductivity from the low-frequency phonons and the «diffusive» modes:

$$\Lambda = \Lambda_{ph} + \Lambda_{dif} \qquad (4a)$$

$$\Lambda_{ph}(T) = 3nk_B v \left(\frac{T}{\Theta_D}\right)^3 \cdot \left[ \int_0^{\Theta_*/T} l(x) \cdot \frac{x^4 e^x}{(e^x - 1)^2} dx \right] \qquad (4b)$$

$$\Lambda_{dif}(T) = 3nk_B v \left(\frac{T}{\Theta_D}\right)^3 \cdot \left[ \int_{\Theta_*/T}^{\Theta_D/T} \alpha \frac{vh}{2k_B xT} \cdot \frac{x^4 e^x}{(e^x - 1)^2} dx \right] \qquad (4c)$$

The results were computer-fitted by the least-square technique to the smoothed thermal conductivity values for sample №1 in the $\alpha$-phase using $n = 2.21 \times 10^{22}$ 1/cm$^3$ and $v = 1.17 \times 10^3$ m/s [2] and varying the parameters C and $\alpha$. The best agreement with experiment was obtained with $C = 3.0 \times 10^{-9}$ cm/K and $\alpha=1.8$. Correspondingly, $C= 0.9 \times 10^{-9}$ cm/K and $\alpha=2.7$ for $CO_2$ and $C=1.5 \times 10^{-9}$ cm/K and $\alpha=2.3$ for $N_2O$ [11]. The fitting to smoothed experimental thermal conductivities and the contributions from the low-frequency phonons $\Lambda_{ph}$ and «diffusive» modes $\Lambda_{dif}$ (calculated by Eqs. (4b-4c)) are shown in Fig. 3.

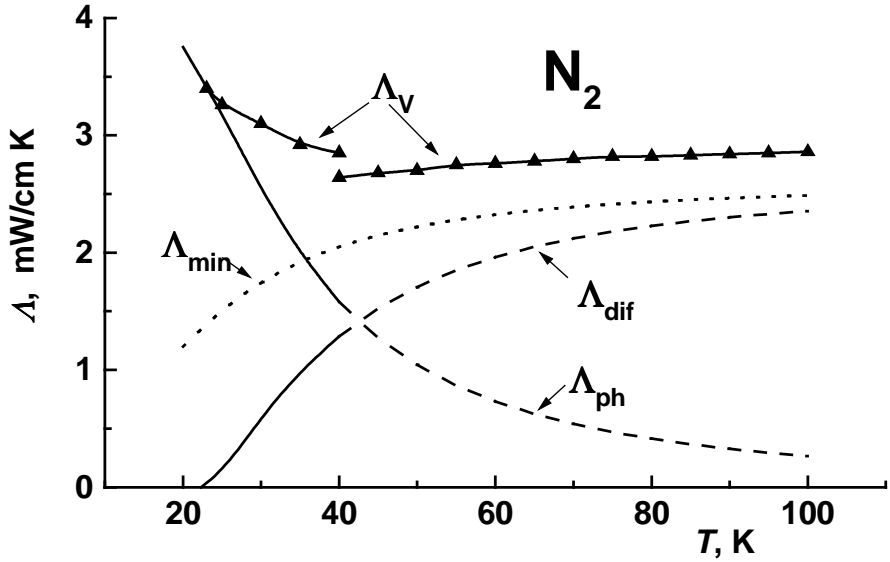

Fig.3. Fitting to smoothed values of experimental thermal conductivity and contributions to thermal conductivity from low-frequency phonons $\Lambda_{ph}$ and «diffusive» modes $\Lambda_{dif}$ calculated according to (4b-4c); (----)- the lower limit of lattice heat conductivity $\Lambda_{min}$ obtained as an asymptote of the dependence $\Lambda_V(T)$.

It is seen that the «diffusive» behavior of the oscillatory modes appears above 20 K, and immediately before the $\alpha \to \beta$ transition nearly half of the heat is transported by the «diffusive» modes. The curves $\Lambda_{ph}$ and $\Lambda_{dif}$ calculated for the $\alpha$-phase of $N_2$ were extrapolated to the region of the existence of the $\beta$- phase. The change from one structure to another may cause a jump of the partial contributions to the thermal conductivity but it will not be too large because a major part of the heat is transported by the «diffusive» modes and they are only slightly sensitive to the structure of the crystal. The lower limit of thermal conductivity $\Lambda'_{min}$ (Fig. 3, broken line) was calculated assuming that all the modes were «diffusive»:

$$\Lambda'_{min} = 3\alpha \left(\frac{\pi}{6}\right)^{1/3} n^{2/3} k_B v \left(\frac{T}{\Theta_D}\right)^2 \int_0^{\Theta_D/T} \frac{x^3 e^x}{(e^x - 1)^2} dx \qquad (5)$$

Note that $\Lambda'_{min}$ is again independent of structure and determined only by the crystal density and hence the Debye temperature were invariant for the constant volume. The lower limit of thermal conductivity $\Lambda'_{min}$ fitted as an asymptote of the dependence $\Lambda_V(T)$ is $\alpha = 1.8$ times higher than the value calculated according to Cahill and Pohl [4]. The discrepancy can partly be accounted for by the imperfection of the model. Nevertheless, there is a certain correlation between $\alpha$ and the number of the degrees of freedom (three translational and z rotational degrees) of molecules: $\alpha \propto (3+z)/3$ [11]. Cahill and Pohl considered amorphous substances and strongly disordered crystals consisting of atoms having no rotational degrees of freedom.

The discussion of the lower limit of thermal conductivity of molecular crystals brings up the inevitable question: Should the site-to-site transport of the rotational energy of the molecules be taken into account? The above correlation suggests the positive answer. In this context, the heat transfer in molecular crystals, solid nitrogen in particular, can be interpreted as follows. At low temperatures, when the phonon and libron branches are well separated, the phonons forming the heat flow are scattered by both phonons and librons [12]. As a result, the thermal resistance increases in comparison with the situation, e. g., in inert gases [3]. As the temperature rises, the phonon-libron interaction enhances and the mixed translation-orientation modes start to transport the heat. The heat transfer increases and extra scattering evolves due to the strong anharmonicity of the librational vibrations. Finally, under a very strong scattering when the heat is transported directly from molecule to molecule (Einstein model), both the rotational and translational energies should equally be taken into account.

In $\beta$-$N_2$ the isochoric thermal conductivity increases slightly with temperature. The absolute value of thermal conductivity is only 10-12% higher than its lower limit $\Lambda'_{min}$. This means that in $\beta$-$N_2$ most of the heat is transported by the «diffusive» modes. The concept of the «lower limit» of thermal conductivity postulates its «saturation» rather than its growth. The increase in the isochoric thermal conductivity with temperature was observed earlier in orientationally disordered phases of some molecular crystals [8]. This effect can be due to the «rotational» component of the total thermal resistance, which decreases as the rotational correlation's between the neighboring molecules become weaker.

The dependence of the thermal conductivity on the molar volume can also be interpreted within this model. The Bridgman coefficient $g = -(\partial \ln \Lambda / \partial \ln V)_T$ is the weighted mean with respect to the phonons and «diffusive» modes whose volume dependences are considerably different [11]:

$$g = \frac{\Lambda_{ph}}{\Lambda} g_{ph} + \frac{\Lambda_{dif}}{\Lambda} g_{dif} \quad , \tag{6}$$

Equation (6) describes the general tendency of the Bridgman coefficient to decrease as more and heat is being transported by «diffusive» modes. The calculation using the procedure [11] and the mean Gruneisen coefficient $\gamma$=2.2 for nitrogen [1,2] gives g = 5.2 at T=35 K and 3.4 at T=60 K, which is in reasonable agreement with the experimental values.

**Conclusions**

The isochoric thermal conductivity of solid $N_2$ has been investigated on four samples of different densities in the temperature interval from 20 K to the onset of melting. In $\alpha$-$N_2$ the isochoric thermal conductivity varies following a dependence weaker than $\Lambda \propto 1/T$; in $\beta$-$N_2$ it increases slightly with temperature. It is shown that the experimental re-sults can be explained within a model in which the heat is transported by low-frequency phonons or by «diffusive» modes above the boundary of mobility. In β-$N_2$ most of the heat transported by the «diffusive» modes. The weak growth of the thermal conductivity in β-$N_2$ can be attributed to the decrease in the «rotational» component of the total thermal resistance due to the relaxing rotational correlations between the neighboring molecules.